\documentclass[aps,prl,twocolumn,showpacs,preprintnumbers,amsmath,amssymb]{revtex4}

\usepackage{graphicx}
\usepackage{dcolumn}
\usepackage{bm}

\begin{document}

\title{Compaction dynamics of metallic nano-foams: A molecular dynamics simulation study}

\author{M. A. Duchaineau, J. B. Elliott, A. V. Hamza, T. Dittrich, T. Diaz de la Rubia and Farid F. Abraham\footnote{To whom queries should be made}}
\affiliation{Lawrence Livermore National Laboratory,Livermore, CA 94550, USA\\}

\date{\today}
\begin{abstract}
We investigate, by molecular dynamics simulation, the generic features associated with the dynamic compaction of metallic nano-foams at very high strain rates. A universal feature of the dynamic compaction process is revealed as composed of two distinct regions: a growing crushed region and a leading fluid precursor. The crushed region has a density lower than the solid material and gradually grows thicker in time by {\it snowplowing}. The trapped fluid precursor is created by ablation and/or melting of the foam filaments and the subsequent confinement of the hot atoms in a region comparable to the filament length of the foam. Quantitative characterization of nano-foam compaction dynamics is presented and the compacted form equation-of-state is discussed. We argue that high-energy foam crushing is not a shock phenomenon even though both share the snowplow feature.
\end{abstract}

\preprint{LLNL-JRNL-468938-DRAFT}
\pacs{52.35.Py, 52.35.Qz}
\maketitle

\section{Introduction}

The emergence of high-energy laser systems, such as the National Ignition Facility (NIF), presents the opportunity to create and interrogate states of matter under unprecedented extreme conditions of pressure, temperature, and strain rates. NIF experiments will also be used to conduct astrophysics and basic science research and to develop carbon-free, limitless fusion energy. In the NIF Energy program, a metallic nano-foam shell has been proposed as an important ingredient in the design of the double-shell Inertial Confinement Fusion (ICF) target. The dynamics of its collapse upon implosion will play an essential role in achieving inertial confinement fusion. However, what is lacking is a fundamental understanding of the relationship between the microphysics of the materials properties of the drive impactor (``reservoir'') and the target materials. An atomic level simulation effort can serve as the basis to relate the microstructure of the reservoir impactor to the pressure profile on the target. The results of these ultra-scale atomic level simulations can be significant in improving the predictive capabilities of continuum level simulations shock compression NIF experiments. 
 
The study of matter under planetary interior conditions of high pressure and relatively low temperature environments is an important research area for NIF \cite{lee-04}. In the traditional laser-based EOS experiment, a strong shock is launched in a material which instantaneously increases both the temperature and pressure tracing the Hugoniot line in the phase diagram. However, broader knowledge of the equation-of-state is required. Recent work by Remington and co-workers \cite{smith-07, lorenz-05, edwards-04} has demonstrated the possibility of a quasi-isentropic compression on a laser based platform. The intense laser pulse drives a reservoir which unloads across a gap onto the sample of interest. Low density porous materials are used as a reservoir because they provide a convenient method to independently vary density without changing the atomic number. The basic idea is to vaporize the foam, and compress the hot vapor, thereby increasing the pressure and temperature of the ``vapor piston'' smoothly. Molecular dynamics simulation techniques can provide design tools to tailor the pressure pulse for isentropic compression of a solid. 
 
In this study, our simulation systems are composite crystal/nano-foam sandwiches. We have constructed physical models and developed the simulation and analysis programs with supporting visualization tools for application to study high strain rate compaction of metallic nano-foam by large-scale computation. Our study incorporates computer-generated nano-foams created by rapidly quenching a high temperature, phase-separating fluid. Our computer foam has been used earlier to study surface-stress-induced relaxation of Au nanostructures  \cite{biener-09}. This present study elucidates the generic features associated with the dynamic compaction of metallic nano-foams at very high strain rates.  Quantitative characterization of nano-foam compaction dynamics is presented and equation-of-state of the compacted form is discussed.

\section{Computer modeling for the simulation}

Our simulation method is Molecular Dynamics (MD), where it is assumed that the motion of the atoms is govern by Newton's second law \cite{abraham-86, allen-87}. The next ingredient is the chosen interatomic potential. It can be simple or complicated. While choosing a complicated potential to describe a particular material is often desired, the physics of a complicated process can be more transparent and discovery of {\it generic} behavior may be more readily forthcoming with the choice of a simple potential.  We assume the Voter-Chen EAM potential for copper as our simple potential \cite{voter-93}. We acknowledge the limitations associated with this choice of potential and will discuss some, such as the neglect of ionization. However, we emphasize that our goal is to elucidate the generic features of the compaction dynamics of nano-foams.

\begin{figure}[ht]
  \includegraphics[width=8.7cm]{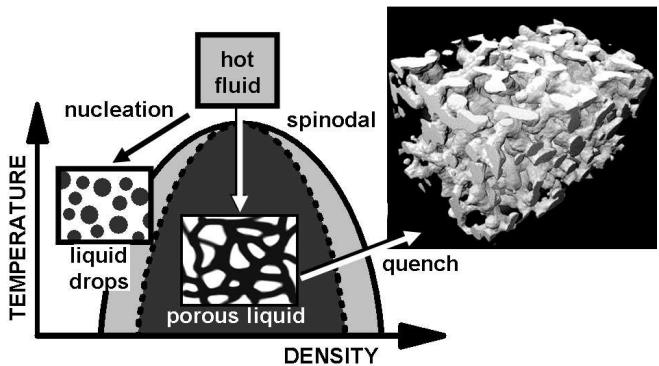}
  \caption{Making computer foam by spinodal decomposition. The method is the rapid quenching of a superheated fluid into the two-phase region followed by the freezing in early time of the phase-separating liquid.}
\label{making-foam}
\end{figure}

Making computer foam was our first step. We took the copper atoms to their super critical state and rapidly quenched it to the middle of its two-phase fluid region of its phase diagram where it underwent phase separation by spinodal decomposition \cite{abraham-79}. See Fig.~\ref{making-foam}. We stopped the fluid phase separation process by rapidly quenching to the solid region, giving rise to a nano-porous foam. The initial porous sample was small and the filaments were amorphous.

For studies of the material response of pore structures under various conditions, it is desirable to construct pore structures at larger length scales, with controlled atomic arrangements (e.g. perfect crystal, grains, etc.), and with specified density profiles or filament cross-sections. Below, we briefly outline a topological filtering approach to analyze and synthesize such designed pore structures using the original small, amorphous sample as a template.

\begin{figure}[ht]
  \includegraphics[width=8.7cm]{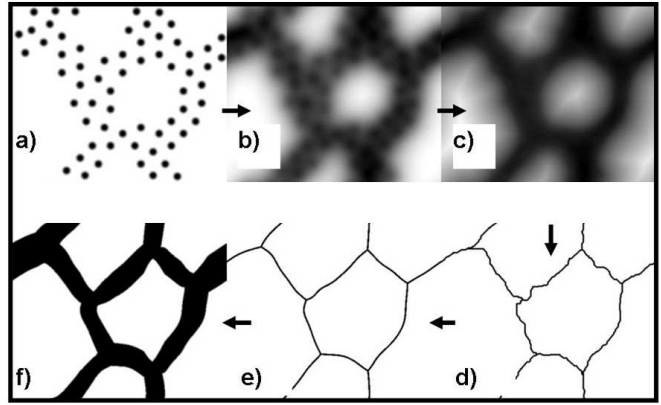}
  \caption{Graphically outlining the topological filtering approach for analyzing and synthesizing nano-pore structures. See text for explanation.}
\label{foam-filaments}
\end{figure}

First, an initial set of atomic positions from the spinodal decomposition process was recorded. Next, a proximity field of original atoms on a regular grid was made. Then a signed distance field relative to the solid/void interface surface was computed. A topology preserving surface propagation was performed from the interface surface to produce a topologically {\it clean} distance field and curved skeleton of the pore filament structure. A distance field was computed from the curved skeleton and then re-scaled to produce uniform density profiles. Finally, the re-scaled distance field was used to {\it carve out} atoms with a pore filament topology that was identical to the original sample, but with specified scales, density profiles and atomic arrangements \cite{gyulassy-07, laney-02}. These steps are illustrated in Fig.~\ref{foam-filaments}. Further details are to be published and a movie of the nano-foam structure may be viewed at http://www.llnl.gov/largevis/atoms/challenge2007 . 

For the three-dimensional pore structure used in this study, a solid/void interface was determined (shown as a green surface in the movie), along with the solid filament centerlines (shown in red in the movie). We have created foams with filament sizes up to 35 nm while experiments are creating foams with filaments of solid copper between 20 and 100 nm in diameter.

\begin{figure}[ht]
  \includegraphics[width=8.7cm]{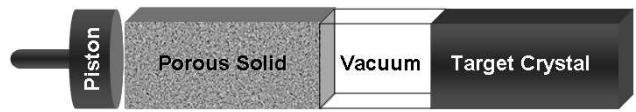}
  \caption{The simulated sandwich configuration. Its building blocks are comprised of foam, vacuum and crystal constructs.}
\label{foam-schematic}
\end{figure}

We used this computer foam to construct the system for the shock simulations. We call it a {\it sandwich}, and its building blocks are comprised of foam, crystal and vacuum constructs. We will study the {\it sandwich configuration} shown in Fig.~\ref{foam-schematic}. 

The purpose of the nano-porous foam block is to provide a source for creating hot vapor by high-energy impact. The anticipated history can be easily summarized. The piston moves at constant velocity to the right, crushing and heating the foam block by a rapidly moving front. We describe this as a {\it compaction} front in contrast to the popular {\it shock} front description. It was anticipated that at sufficiently high piston velocities, the foam completely vaporizes. Hot vapor expands from the heated foam into the vacuum and is compressed by the driving piston. This, in turn, creates a continually rising pressure on the face of the right-hand target crystal. In this paper, our primary interest is the qualitative features of the equation of state of the compacting foam. The initial density of the computer foam in this study is taken to be 15\% of the density of solid copper.

\section{The simulations}

\subsection{Universal dynamical features of a shocked sandwich}

\begin{figure}[ht]
  \includegraphics[width=8.7cm]{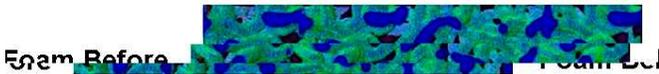}
  \caption{Top: the nano-foam prior to impact. Bottom: the universal feature of the compaction process: the compacted foam (at left) and the vapor precursor (at right). The colors represent the potential energy of the individual atoms. The limits are blue being the energy of a solid copper atom and red being a weakly bonded or isolated atom.}
\label{impacted-filaments}
\end{figure}

A universal feature of the compaction process traveling through the porous solid has emerged from our simulations. The feature has two regions: the {\it growing compacted region} and the {\it trapped fluid precursor}. See Fig.~\ref{impacted-filaments}.

\begin{figure}[ht]
  \includegraphics[width=8.7cm]{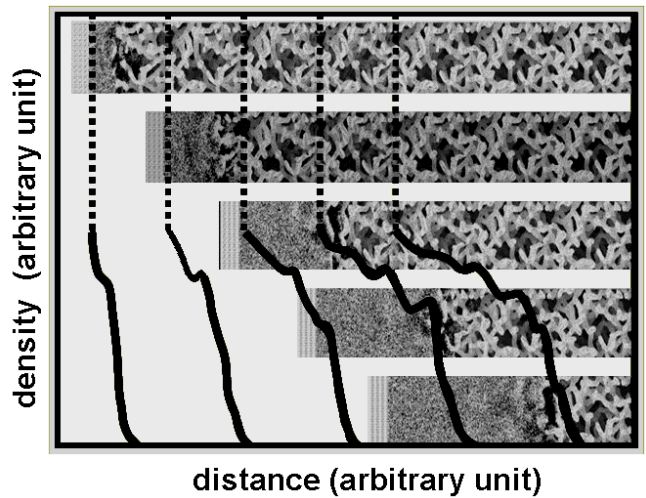}
  \caption{Growing densification of the crushed foam as a consequence of snowplowing. The vertical dash lines define the piston position and the solid lines are the respective instantaneous density profiles. Superimposed in the background are snapshots of the sandwich.}
\label{foam-density}
\end{figure}

In contrast to a shocked solid which achieves constant density immediate behind a sharp front, foam compaction is an evolving phenomenon where growing densification occurs as the region falls back from the forward moving broad front as shown in Fig.~\ref{foam-density}. The compacted region has a peak density lower than its cold solid material. A feature shared by shocking and by compacting is that growth is achieved by snowplowing. 

\begin{figure}[ht]
  \includegraphics[width=8.7cm]{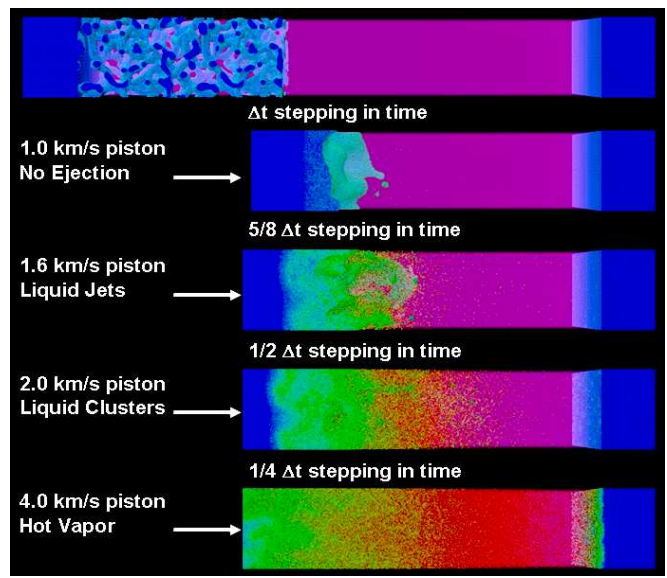}
  \caption{The compaction precursor breaking out into the vacuum. This fluid precursor can be liquid or vapor, depending on the piston impact speed. The individual snapshot times are normalized so that the apparent rates of collapse of the four sandwiches are the same. See caption of Fig.\ref{foam-density} for description of colors.}
\label{foam-compaction}
\end{figure}

A trapped fluid precursor is created by ablation and/or melting of the foam filaments and the subsequent trapping of the hot atoms within a {\it mean-free-path} that is comparable to the filament length of the foam. The high-energy fluid atoms do not experience free expansion. Instead, they traverse a maze of uncorrelated, random channels restricting free flow to the vacuum. This fluid precursor can be liquid or vapor, depending on the piston impact speed. See Fig.~\ref{foam-compaction}. The fluid precursor rapidly achieves an approximately fixed length, traveling at a speed comparable to the compaction front, $U_f$. The compaction front speed is greater than the piston speed, $U_p$, but slower than the shock speed in the perfect solid. This picture is consistent with a recent experiment where a calorimeter measurement suggested that a significant temperature rise preceded the compaction front traveling through a foam \cite{dittrich-09}. 

\subsection{Features of the ejecta leaving the foam}

\begin{figure}[ht]
  \includegraphics[width=8.7cm]{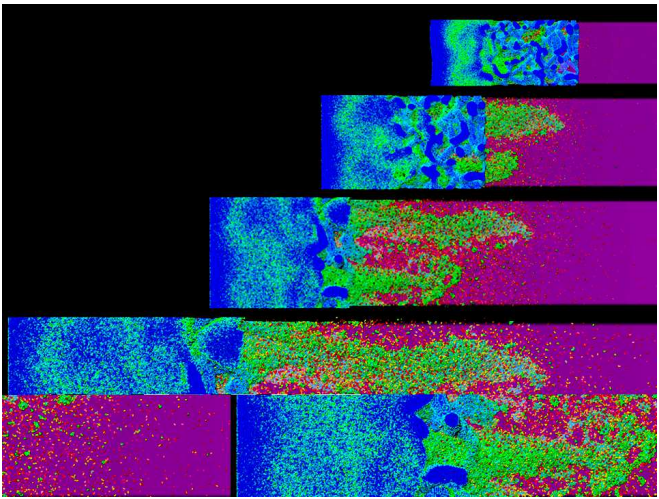}
  \caption{Snapshots of the liquid ejecta at $U_p = 1.6$ km/s. See caption of Fig.\ref{foam-density} for description of colors.}
\label{liquid-ejecta}
\end{figure}

We drove our sandwich with pistons that travelled at speeds of interest to recent experiments and discovered a {\it phase change} associated with the ejecta at the slower speeds. This is best seen as the ejecta leaves the foam boundary as shown in Figures~\ref{foam-compaction} and \ref{liquid-ejecta}.

For $U_p \le 1$ km/s, there was only foam compaction and no ejecta. As $U_p$ was increased from 1 km/s to 4 km/s, we observed a liquid front, passing to liquid jetting, liquid cluster spraying, and, at the highest speed, hot vapor ejecta. Since the recent EOS experiments are presently approaching $U_p \sim 20$ km/s, we can conclude that the ejecta will be a hot vapor at these higher piston speeds.

\subsection{EOS simulations of the compacting foam}

\begin{figure}[ht]
  \includegraphics[width=8.7cm]{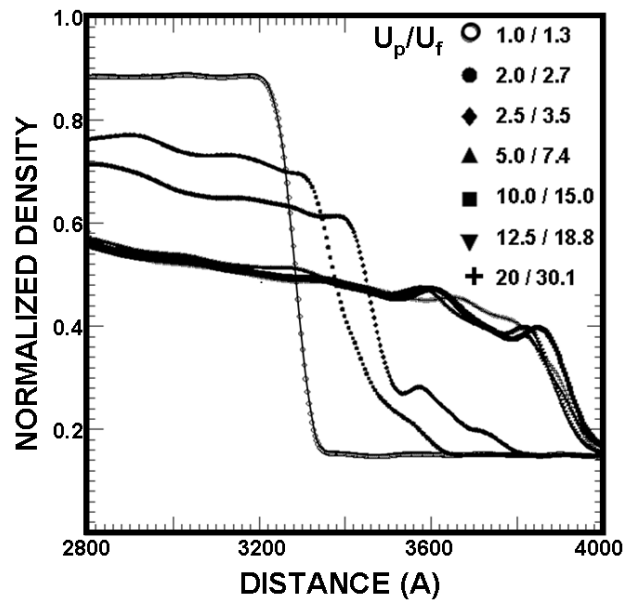}
  \caption{Density profiles for piston speeds from 1 to 20 km/s, respectively. The snapshot times are chosen so as to capture the same piston location near breakout.}
\label{density-profile}
\end{figure}

Figure~\ref{density-profile} shows the respective density profiles for piston speeds from 1 to 20 km/s at decreasing times so as to capture the same piston location. The chosen piston location is where the compacted regions are approaching their greatest thickness; i.e., the respective fronts are near breakout or the foam-vacuum interface.

\begin{figure}[ht]
  \includegraphics[width=8.7cm]{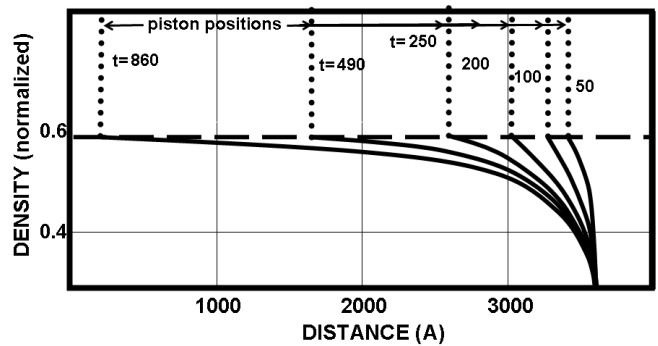}
  \caption{Smoothed density profiles at increasing time $t$ (in arbitrary units) for $U_p = 5$ km/s simulation. The profiles are centered with respect to the compaction front position and smoothed to demonstrate the form change. The dotted vertical lines show the position of the piston.}
\label{norm-density}
\end{figure}

Figure~\ref{density-profile} also shows that the density profiles are nearly identical for $U_p \ge 5$ km/s. There is not perfect overlap of the density profiles in the region around $3,500$ A, but the variation is not correlated with $U_p$. The commonality of the different profiles is striking and quite unexpected.

Extending the $U_p = 5$ km/s simulation to four times in time and distance, we note that the smoothed profile of the compacted foam maintains its shallow slope with the contact density at the piston position not changing. This is shown in Fig.~\ref{norm-density}.

\begin{figure}[ht]
  \includegraphics[width=8.7cm]{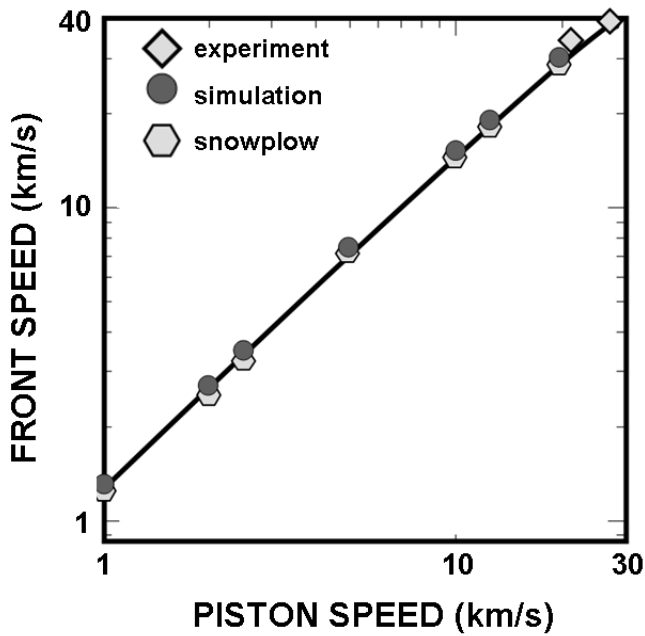}
  \caption{Comparison of the front speeds $U_f$ as a function of the piston speeds $U_p$ for the simulations. The diamonds show results of experiments \cite{page-09}.  The circles show the results of MD simulations.  The hexagons show estimates based on the Eq.~(\ref{mass-cons}).}
\label{Uf-Up}
\end{figure}

Mass and momentum conservation applied via the simple {\it snowplow model} can explain the foam densification of the compaction front \cite{zeldovich-02}. Conservation of mass is the essential ingredient for predicting the front speed and yields the simple relation 
\begin{equation}
\frac{U_f}{U_f} = \left( 1 - \frac{\rho_0}{\rho_c} \right) ^{-1}
\label{mass-cons}
\end{equation}
where $\rho_0$ and $\rho_c$ are the non-compacted foam density and the compacted foam density. Equation~(\ref{mass-cons}) has a weak dependence on the $\rho_0$ and $\rho_c$ and, as such, does not place a high demand on modeling accuracy. Thus one should not be surprised at the agreement between our simple model (that assumes {\it billiard ball} atoms that remain neutral irrespective of temperature and pressure) and the very recent (but limited) experiment data \cite{page-09} shown in Fig.~\ref{Uf-Up}.

Also evident in Fig.~\ref{Uf-Up} is good agreement between the simulated speeds and the snowplow model; {\it i.e.}, the jump condition prediction of Eq.~(\ref{mass-cons}). The compacted regions do not have a strictly flat (constant) density, but assuming an average density is sufficient to obtain excellent agreement. While this agreement is not a sensitive measure of our knowledge of the compaction dynamics, this insensivity plays an important role in estimating ionization effects discussed later in the paper.

\begin{figure}[ht]
  \includegraphics[width=8.7cm]{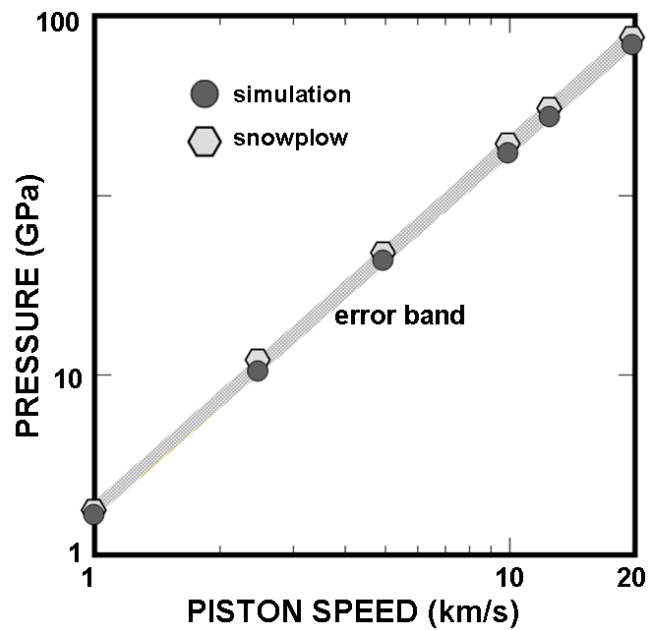}
  \caption{Comparison of pressures as a function of piston speeds for the simulations and the snowplow jump predictions. The error band is estimated using the simulation uncertainly in the virial pressure.}
\label{P-Up}
\end{figure}

Similarly, the atomic virial pressure of Kirkwood and the snowplow estimate,
\begin{equation}
P = \rho_0 U_f U_p
\label{pressure}
\end{equation}
are in agreement. See Fig.~\ref{P-Up}. This is a consequence of momentum conservation across the front, a global conservation requirement that is necessarily satisfied. The snowplow pressure $P$ is calculated using the measured front speed from simulation. Again, we realize that such agreement is not a sensitive measure of our knowledge of the compaction EOS by simply examining Eq.~(\ref{pressure}).

We mention an important point in calculating the pressure from the virial equation. Care must be taken to use the local density in this expression for calculating the pressure profile. Using the local density, one gets the correct virial pressure which is constant throughout the compacted foam (even though the density is not) and equal to the pressure estimated from the conservation of momentum requirement. 

\begin{figure}[ht]
  \includegraphics[width=8.7cm]{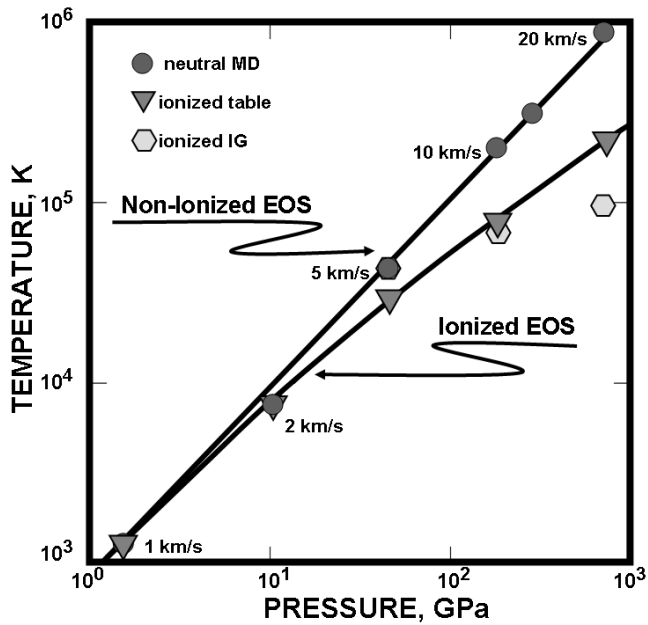}
  \caption{Equation of state of compacted copper foam. The triangles show results from the MD simulations.  Downward pointing triangles show results using the tables (see text for details). hexagons show results from an ideal gas approximation (see text for details). The curves are the results of fits to the points and are included to guide the eye.}
\label{T-P}
\end{figure}

The equation of state EOS of the compacted copper foam is presented in Fig.\ref{T-P}. The figure shows both {\it non-ionizing} and ionizing states. The applicability of the {\it non-ionizing} EOS for real copper foam is in doubt much beyond $U_p = 10$ km/s because of ionization of the copper atoms. To estimate the effect of ionization on the EOS, we consider two different methods. The first method uses tabulated EOS for hot-dense matter based on sophisticated model calculations and experiment \cite{more-88, rozsnyai-01}.

In our earlier discussion of the snowplow model, we emphasized that global conservation of mass and momentum leading to Equations~(\ref{mass-cons}) and (\ref{pressure}) demonstrate a weak density dependence of front speed, as well as the pressure. Furthermore, the front speed found from simulation agrees with the experimental measurements. This leads us to assume that the density and pressure for a given simulation is the same for both the neutral and ionized state, and we employed these tables to determine the temperatures of the ionized state. The results are shown in Fig.~\ref{T-P}. We note that the ionized state temperature at $U_p = 20$ km/s is a factor of four lower than the MD result.

\begin{figure}[ht]
  \includegraphics[width=8.7cm]{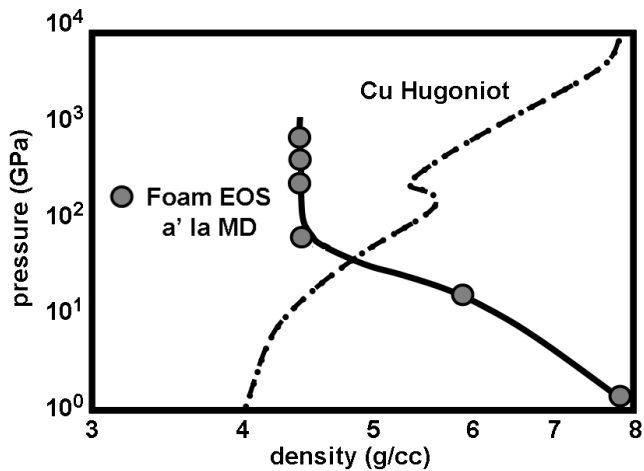}
  \caption{This studyÕs pressure-density EOS compared to the known copper Hugoniot EOS. The solid line is drawn to guide the eye.}
\label{hugoniot}
\end{figure}

For the second method, we invoke conservation of energy in the ideal gas (IG) approximation \cite{zeldovich-02}. We discuss this IG approach for pedagogical reasons even though the table lookup is more accurate. We denote as state 0, the crushed foam where the copper atoms are constrained to be neutral. This is our MD case. Energy, number of atoms and average number of electrons ionized per atom are denoted by $E$, $N$ and $Z$, respectively. In the ideal gas approximation, the energy is 
\begin{equation}
E_0 = \frac{3}{2} N T_0
\label{energy}
\end{equation}
for the neutral system and
\begin{equation}
E_i = \frac{3}{2} N \left( 1 + Z_i \right) T_i + N e_i
\label{ion-energy}
\end{equation}
for the ionized system where $e_i$ average energy used to free $Z_i$ electrons from an atom. Assuming the energy and the number of atoms are conserved, we find then the following relation:
\begin{equation}
T_0 = \left( 1 + Z_i \right) + \frac{2}{3}e_i
\label{ion-temp}
\end{equation}
The ionization energies of copper are 0 to +1: 745.5 kJ/mol, +1 to +2: 1958 kJ/mol, +2 to +3: 3554 kJ/mol, +3 to +4: 5326 kJ/mol. Using the Eq~(\ref{ion-temp}) between the two temperatures, we go up the ionization scale until it predict the lowest positive ionized temperature. In Fig.~\ref{T-P}, the ideal gas (IG) predictions are plotted as hexagons. At the highest pressure, the IG ionization lowers the temperature by an order-of-magnitude. This is a very simplistic model is pedagogical and shows explicitly why there would be a lower temperature due to ionization.

\begin{figure}[ht]
  \includegraphics[width=8.7cm]{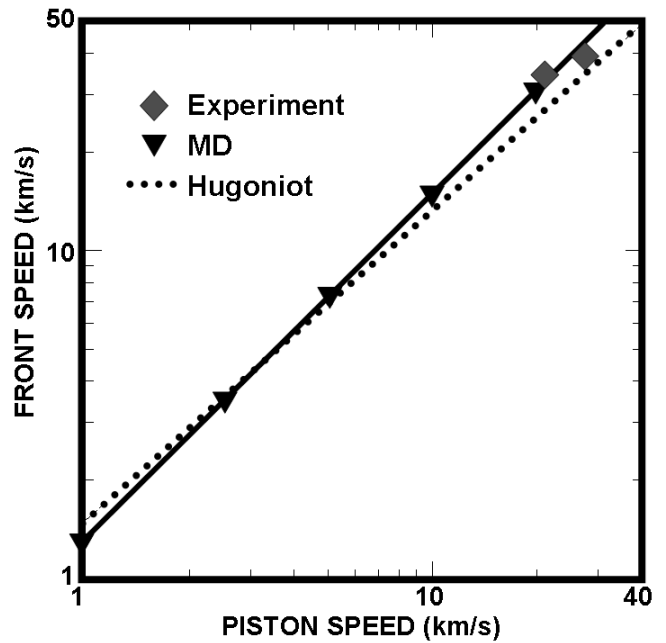}
  \caption{The speed of the compaction front as a function of the speed of the piston. Diamonds show experimental results \cite{page-09}. Downward pointing triangles show results of the MD simulation; the solid line is a fit to the MD results. The dotted line shows the theoretical copper Hugoniot \cite{more-88, rozsnyai-01}.}
\label{Uf-Up-2}
\end{figure}

Figure~\ref{hugoniot} shows the pressure-density EOS from the MD simulations compared to the theoretical copper Hugoniot \cite{more-88, rozsnyai-01}. We note that the compacted foam state is not accurately represented by the shock hugoniot of low density copper foam. This is not surprising from all that we have learned earlier. One must be cautious in attempting to characterize the high-energy crushing of foam as a shock phenomenon. However, this does not preclude the very good agreement of $U_f(U_p)$ between the MD simulations and the copper foam hugoniot.  Comparison of $U_f(U_p)$ for MD, experiment and the copper foam hugoniot is shown in Fig.~\ref{Uf-Up-2}. The agreement between all three is good, though MD and experiment is best.

\section{Summary}

This study was driven by the NIF Energy and Science programs. We investigated the generic features of the compaction dynamics of metallic nano-foams as related to current ICF and science interests. A successful atomistic method was implemented for creating a {\it computer} nano-foam. The microscopic processes in the compaction dynamics showed that a strong compaction front passing through a foam induces acceleration, crushing, ablation and mixing of the foam material within the porous background. Depending on the piston speed, the ejecta could be a solid front, a liquid, an imperfect vapor, or a hot gas. The MD simulations agreed with limited experimental measurements. As should be expected, the simple snowplow model gave an overall accurate macroscopic description of the dynamics. The EOS of the compacted foam is a direct consequence of the atomistic simulation, though its accuracy depends of a robust description of the atomic behavior over a wide range temperatures and pressures. Our simple interatomic potential did not fulfill that requirement because of the constrained neutrality of the atoms. However, simple model estimates of the ionization were made and are believed to be reasonable. The crushing of the foam is a highly complex evolving state as one moves from the front toward the piston.

\section{Acknowledgments}

This work performed under the auspices of the U.S. Department of Energy by Lawrence Livermore National Laboratory under Contract DE-AC52-07NA27344.


\begin{thebibliography}{99}

\bibitem{lee-04}
	R. W. Lee, D. Kalantar and J. Molitoris, UCRL-TR-203844  (2004).
	
\bibitem{smith-07}
	 R. F. Smith, K. T. Lorenz , D. Ho, B. A. Remington, A. Hamza, J. Rogers, S.  Pollaine, S. Jeon, Y.-K. Nam and J. Kilkenny,  Astrophysics and Space Science {\bf 307}, 269 (2007).

\bibitem{lorenz-05}
	K. T. Lorenz, M. J. Edwards, S. G. Glendinning, A. F. Jankowski, J. McNaney, S. M.  Pollaine and  B. A. Remington, Physics of Plasmas {\bf 12}, 056309 (2005).
	
\bibitem{edwards-04}
	J. Edwards, K. T. Lorenz, B. A. Remington, S. Pollaine, J. Colvin, D. Braun, B. F. Lasinski, D. Reisman,  J. M. McNaney and J. A. Greenough, R. Wallace, H. Louis, D. Kalantar,  Physical Review Letters {\bf 92}, 075002, (2004). 
	
\bibitem{biener-09}
	J. Biener,  A. Wittstock, L. Zepeda, M. Biener, V. Zielasek, D. Kramer, R. Viswanath, J. Weissmuller, M. Baumer and A. Hamza,  Nature Materials {\bf 8}, 47 (2009).

\bibitem{abraham-86}
	 F. F. Abraham, Advances in Physics {\bf 35}, 1 (1986).
	
\bibitem{allen-87}
	M. P. Allen and  D. J. Tildesley, {\it Computer Simulation of Liquids} (Clarendon Press, Oxford, 1987).
		
\bibitem{voter-93}
	A. F. Voter, Technical Report LA-UR 93-3901, Los Alamos National Laboratory (1993).

\bibitem{abraham-79}
	F. F. Abraham, Journal of Chemical Physics {\bf 70}, 2577 (1979).
	
\bibitem{gyulassy-07}
	A. Gyulassy, M. Duchaineau, V. Natarajan, V. Pascucci, A. Higginbotham and B. Hamann, IEEE Transactions on Visualization and Computer Graphics {\bf 13}, 1432 (2007).
	
\bibitem{laney-02}
	 D. Laney, M. Bertram, M. A. Duchaineau and N. Max, Proceedings of the First International Symposium on 3D Data Processing, Visualization, and Transmission, Padova Italy, June 19-21, 470 (2002).
 
\bibitem{dittrich-09}
	T. Dittrich, private communication (2009).

\bibitem{zeldovich-02}
	Ya. B. ZelÕdovich and r Yu. P. Raize, {\it Physics of Shock Waves and High Temperature Hydrodynamic Phenomena} (Dover, New York, 2002).

\bibitem{page-09}
	R. H.  Page, private communication (2009).
 
\bibitem{more-88}
	R. M.More, K. H. Warren, D. A. Young and G. B. Zimmerman, Phys. Fluids {\bf 31}, 3059 (1988).

\bibitem{rozsnyai-01}
	B. F. Rozsnyai, J. R. Albritton,  D. A. Young, V. N. Sonnad and D. A. Liberman, Phys. Lett. {\bf A 291}, 226 (2001).
 
\end{thebibliography}
\end{document}